\documentclass{ws-p9-75x6-50}
\usepackage{color}
\raggedright
\usepackage{psfig}
\def\lesssim{\mathrel{\mathchoice {\vcenter{\offinterlineskip\halign{\hfil
$\displaystyle##$\hfil\cr<\cr\sim\cr}}}
{\vcenter{\offinterlineskip\halign{\hfil$\textstyle##$\hfil\cr<\cr\sim\cr}}}
{\vcenter{\offinterlineskip\halign{\hfil$\scriptstyle##$\hfil\cr<\cr\sim\cr}}}
{\vcenter{\offinterlineskip\halign{\hfil$\scriptscriptstyle##$\hfil\cr<\cr\sim\cr}}}}}
\def\grsim{\mathrel{\mathchoice {\vcenter{\offinterlineskip\halign{\hfil
$\displaystyle##$\hfil\cr>\cr\sim\cr}}}
{\vcenter{\offinterlineskip\halign{\hfil$\textstyle##$\hfil\cr>\cr\sim\cr}}}
{\vcenter{\offinterlineskip\halign{\hfil$\scriptstyle##$\hfil\cr>\cr\sim\cr}}}
{\vcenter{\offinterlineskip\halign{\hfil$\scriptscriptstyle##$\hfil\cr>\cr\sim\cr}}}}}
\def\sRR{{\sl \hbox{I\kern-.2em\hbox{R}}}} 

\def\cblue{\color{blue}}
\def\cb{\color{black}}

\def\cred{\color{red}}

\definecolor{dm}{rgb}{.5,0,.5}
\def\cdm{\color{dm}}
\def\apj{Astrophys. J.}
\def\prl{Phys. Rev. Lett.}
\def\prd{Phys. Rev. D }
\begin{document}

\title{Implications of the r-Mode Instability of Rotating Relativistic Stars}

\author{John L. Friedman}

\address{Department of Physics, University of Wisconsin-Milwaukee, 
PO Box 413, Milwaukee, WI 53201}

\author{Keith H. Lockitch}

\address{Center for Gravitational Physics and Geometry, Department of 
Physics, \\ Pennsylvania State University, 104 Davey Laboratory,
State College, PA 16802}

\maketitle

\abstracts{
Several recent surprises appear dramatically to have improved the
likelihood that the spin of rapidly rotating, newly formed neutron
stars (and, possibly, of old stars spun up by accretion) is limited by
a nonaxisymmetric instability driven by gravitational waves.  Except
for the earliest part of the spin-down, the axial l=m=2 mode (an
r-mode) dominates the instability, and the emitted waves may be
observable by detectors with the sensitivity of LIGO II.  
A review of these hopeful results is followed by a discussion of 
constraints on the instability set by dissipative mechanisms,
including viscosity, nonlinear saturation, and energy loss to a
magnetic field driven by differential rotation.  
}

\section{Introduction}
\label{sec:intro}

We review here recent work on a gravitational-wave driven instability
that may sharply limit the spin of young, rapidly rotating neutron
stars.  Andersson and Kokkotas\cite{akrev} have given a comprehensive
review of recent work on this r-mode instability and its physical
implications, following several earlier reviews\cite{flrev,lindrev,s98}.  
The last of these, a general review of the
gravitational-radiation driven instability, is more pedagogical and not
limited to work on r-modes.  Included in the present review is a survey
of recent work on nonlinear saturation and mode coupling not included
in these earlier articles.

The first of the surprises mentioned in the abstract was the discovery
that the r-modes, rotationally restored modes that have axial parity
for spherical models, are unstable in perfect fluid models with
arbitrarily slow rotation.  Indicated in numerical work by Andersson~\cite{a97}, 
the instability is implied in a nearly Newtonian context by
the Newtonian expression for the r-mode frequency~\cite{pp78,a97,fm98}, and
a computation by Friedman and Morsink~\cite{fm98} of the canonical
energy of initial data showed (independent of assumptions about the
existence of discrete modes) that the instability is a generic feature
of axial-parity perturbations of relativistic stars.

Studies of the viscous and radiative timescales associated with the
r-modes (Lindblom et al.\cite{lom98}, Owen et al.~\cite{o98},
Andersson et al.~\cite{aks98}, Kokkotas and Stergioulas~\cite{ks98},
Lindblom et al.~\cite{lmo99}) revealed a second surprising
result:  The growth time of r-modes driven by current-multipole
gravitational radiation is significantly shorter than had been
expected, so short, in fact, that the instability to gravitational
radiation reaction easily dominates viscous damping in hot, newly
formed neutron stars (see Fig. \ref{fig:owen} below).  As a result, a
neutron star that is rapidly rotating at birth now appears likely to
spin down by radiating most of its angular momentum in gravitational
waves. (See, however, the caveats below.)

Nearly simultaneous with these theoretical surprises was the discovery
by Marshall et.al.~\cite{m98} of a fast (16ms) pulsar in a supernova
remnant (N157B) in the Large Magellanic Cloud.  From the pulsar's
period and period derivative, and from the estimated age of the
remnant, the initial period is estimated at less than 10ms, implying a
class of neutron stars that are rapidly rotating at birth.  Fortifying
this conclusion is the belief that accretion-induced collapse of
O-Ne-Mg and C-O dwarfs leads to rapidly rotating neutron stars~\cite{nk91}
and the likelihood that magnetars are rapidly rotating at birth.  

Spurred by these surprises, over fifty authors have worked on aspects
of the r-mode instability:
\begin{itemize}  
\item Examining the modes themselves, for Newtonian and relativistic
models.~\cite{a97,a98,bk99,kye00,kyye00,k98,kh99,kh00,ks98,li98,lmo99,lock99,lf99,laf00,laf01,rk01,y01,ye97,yl00a,yl00b,ykye00}
\item Studying the waveforms and detectability of the gravitational waves 
they emit~\cite{aks98,akst98,bc98,fms98,h98,l99,lom98,mad98,mad00,o98,rjm99,sfm99}
\item Finding the dominant mechanisms of the effective shear and bulk 
viscosities, including effects of a superfluid interior and a solid
crust.~\cite{ac00,ajks00,aks98,bu00,l98,lu00,lm99,lou00,m01,rm00,r00,wma00,yl01}
\item Finding the maximum mode amplitude permitted by nonlinear fluid 
evolution.~\cite{fls01,lu99,ltv00,saftw01,sf00}
\item Finding the nonlinear differential rotation associated with r-modes 
and asking whether such rotation dissipates in a magnetic field the energy 
of an unstable r-mode.~\cite{fls01,hl00,lu99,rls99,s99}
\end{itemize}

\section{The Gravitational-Wave Driven Instability}

All rotating perfect fluid stars are subject to a nonaxisymmetric
instability driven by gravitational radiation.  The instability was
found by Chandrasekhar~\cite{ch70} for the $\cblue l=m=2$ polar mode of
the uniform-density, uniformly rotating Maclaurin spheroids.  Although
this mode is unstable only for rapidly rotating models, by looking at
the canonical energy of initial data with arbitrary values of $\cblue
m$, Friedman and Schutz~\cite{fs78b} and Friedman~\cite{f78} showed
that the instability is a generic feature of rotating perfect fluid
stars, that even slowly rotating perfect-fluid models are formally
unstable.

For a normal mode of the form 
$\cblue e^{i(\sigma t+m\varphi)}$ this nonaxisymmetric instability 
acts in the following manner:  
In a non-rotating star, gravitational radiation removes positive 
angular momentum from a  forward moving mode and  negative angular 
momentum from a backward moving mode, thereby damping all 
time-dependent, non-axisymmetric modes. In a star rotating 
sufficiently fast, however, \color{dm} a backward moving mode can be dragged 
forward \cb as seen by an inertial observer; and it will then radiate
positive  angular momentum.  The angular momentum of the mode, however,
remains negative, because the perturbed star has lower total angular 
momentum than the unperturbed star.  As positive angular momentum
is removed from a mode with negative angular momentum, the angular 
momentum of the mode becomes increasingly negative, implying that
its amplitude increases: The mode is driven by gravitational radiation.

The conclusion, that a mode is unstable if it is prograde relative to 
infinity and retrograde relative to the star is equivalent to requiring
that its frequency satisfies the condition,
\begin{equation}\color{dm}
\sigma(\sigma+m\Omega) < 0.
\label{criterion}
\end{equation}
For the polar f- and p-modes, the frequency is large and 
approximately real.  Condition (\ref{criterion}) will be met 
only if $\cblue|m\Omega|$ is of order $\cblue |\sigma|$, so that for a given 
angular velocity the instability will set in first through modes 
with large $\cblue m$. 

\section{Unstable r-Modes}
\label{sec:mode_character}

Part of the reason that axial modes (r-modes) were not studied
extensively for neutron stars is that for spherical stars they are
stationary, convective currents. \footnote{This statement, although
formally true, is somewhat misleading.  A solution with axial parity to
the linearized Euler equation {\em is} a stationary current, but the
nonlinear terms in the Euler equation are important once a fluid
element has moved a distance of order the radius of the star, and the
nonlinear solution is time-dependent.}  That is, axial perturbations 
belonging to an $l,m$ representation of the rotation group behave under 
parity as $(-1)^{l+1}$, opposite to $Y_{lm}$.  Because any scalar can 
be written as a sum of $Y_{lm}$, no scalar perturbation of a spherical star 
has axial parity. Axial perturbations have, to order $\Omega^2$, vanishing 
$\delta p, \delta \rho$ and $\delta \Phi$. Only their velocity perturbation
is nonzero, and it has the form
\be\cblue
A r^l {\bf r}\times \nabla Y_{lm}.
\ee
A diagram of the velocity field (due to L. Lindblom) for $l=m=2$
is reproduced below.  

\begin{figure}
\centerline{\psfig{file=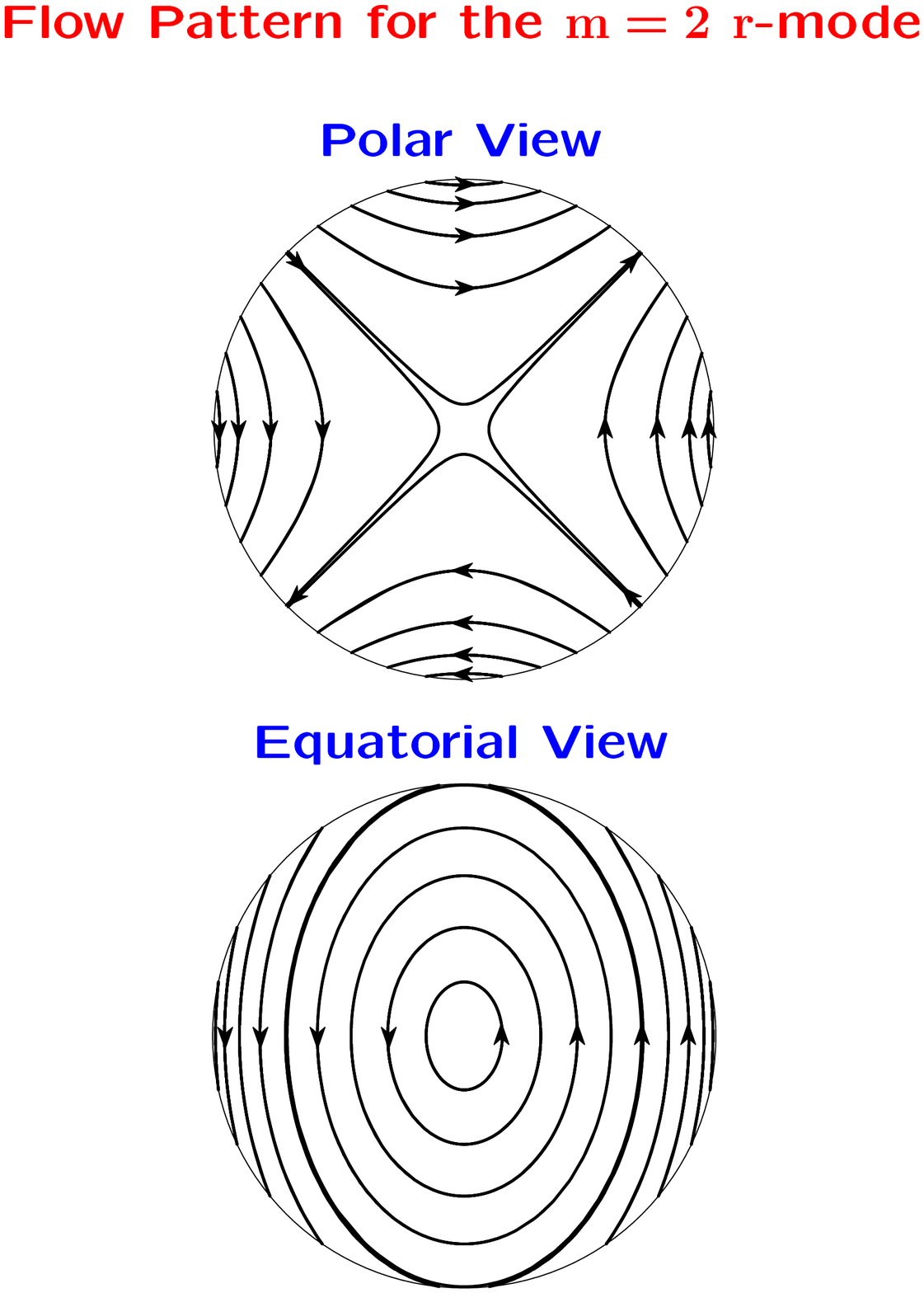,height=6in}}
\caption{The perturbed velocity field of the $\cblue l=m=2$ r-mode.}
\label{fig:rmode1}
\end{figure}

\begin{figure}
\epsfxsize=13pc % will enlarge or reduce the postscript figures based on the xsize
\epsfbox{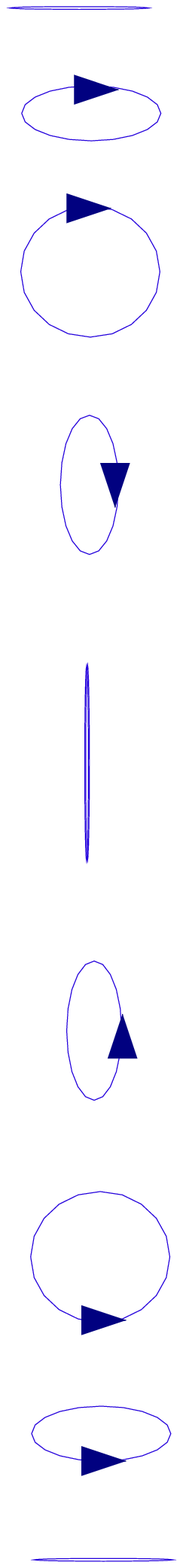} % postscript image file name
\caption{Fluid trajectories of the l=m=2 r-mode}  
\label{fig:rmode2}
\end{figure}

The fluid elements move along the integral curves of $\cblue\bf v+\delta v$, 
ellipses, to first order in the perturbation, as shown in Fig.~\ref{fig:rmode2}.
The restoring force in a rotating frame can be regarded as the Coriolis 
force, leading to a frequency of oscillation proportional to angular 
velocity of the star.  

At the onset of instability, a mode's frequency vanishes.  For f-modes
and p-modes (both polar modes) with low values of $\cblue l$, the frequency of
a mode is of order $\cblue \sigma \sim 1/t_{\rm dynamical} \sim \sqrt{G\rho}$,
implying instability points at similarly large values of the star's
angular velocity.\cite{msb,s98}  That is, to drag a mode that is retrograde
relative to the fluid forward relative to infinity, one must have
$\cblue \Omega \sim \sigma_{\rm spherical star}/l$.  In fact, when viscosity is 
taken into account, polar modes are estimated to be unstable in a uniformly
rotating star only for $\cblue \Omega > 0.9 \Omega_K$, where $\cblue \Omega_K$ is the
Kepler limit on a star's rotation, the angular velocity of a satellite
at its equator.~\cite{il91,cl92,l95,lm95} \footnote{This estimate may 
decrease, perhaps to {\cblue 0.8} $\cblue \Omega_K$ when a 
general-relativistic computation with viscosity is carried out.}

The frequency of an r-mode, however is already zero for a spherical 
star; and {\it\cdm for any nonzero rotation, every r-mode of a perfect-fluid 
model is unstable}.\cite{a97,fm98} The instability
in models of slowly rotating, nearly Newtonian stars, follows from the
fact that the frequencies~\cite{pp78} satisfy the criterion 
(\ref{criterion}), 
\begin{equation}\cblue 
\sigma(\sigma+m\Omega) = 
- \frac{2(l-1)(l+2)m^2\Omega^2}{l^2(l+1)^2} < 0.
\end{equation}
As noted below, it is likely, but not yet certain that relativistic
stars generically have a discrete spectrum of r-modes; but a computation by
Friedman and Morsink~\cite{fm98} of the canonical energy of initial
data is independent of the existence of normal modes, and it shows that the
instability is a generic feature of axial-parity fluid perturbations of
relativistic stars.

For stars with general equations of state, the r-modes 
describe the dynamical evolution of initial perturbations that have
axial parity.\cite{pp78,s82,sm83} In barotropic Newtonian models,
however, the only purely axial modes allowed are the r-modes with $\cblue l=m$
and simplest radial behavior.\cite{pea81} The \cdm
disappearance of the purely axial modes with $\cdm l\neq m$ \cb occurs
for the following reason.\cite{lf99}  Axial perturbations of a
spherical star are time-independent convective currents with vanishing
perturbed pressure and density.  In spherical {\sl barotropic} stars,
stars for which both star and perturbation are governed by a single
one-parameter equation of state, the gravitational restoring forces
that give rise to the g-modes vanish, and they, too, become
time-independent convective currents with vanishing perturbed pressure
and density.  Thus, the space of zero frequency modes, which generally
consists only of the axial r-modes, expands for spherical barotropic
stars to include the polar g-modes.  This large degenerate subspace of
zero-frequency modes is split by rotation to zeroth order in the star's
angular velocity, and the corresponding modes of rotating barotropic
stars are generically hybrids whose spherical limits are hybrids of
axial and polar perturbations.  Because their restoring force is rotational
(Coriolis), we have referred both to them and to the r-modes as rotational 
modes, and they are called ``inertial modes'' in the fluid-dynamics 
literature.\cite{g64}

\begin{figure}
\centerline{\psfig{file=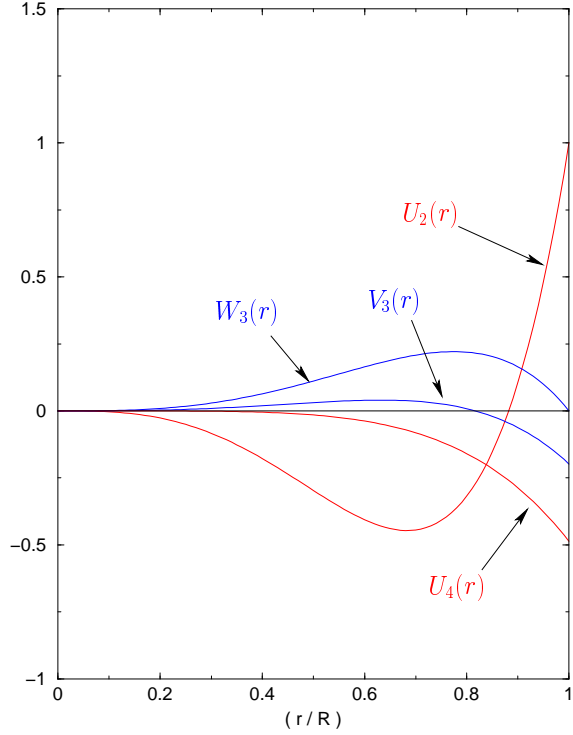,height=4in}}
\caption{An r-g ``hybrid'' rotational mode of a uniform density
Newtonian star. The functions shown are coefficients of the
spherical harmonic expansion (\ref{v_exp}) of the perturbed fluid 
velocity.}
\label{fig:newt}
\end{figure}

To compute these ``hybrid'' rotational modes, Lockitch and Friedman 
\cite{lf99} expand the perturbed 3-velocity, $\delta v^a$, in vector 
spherical harmonics (see also Ref. \cite{lsvh92}),
\begin{equation}
\delta v^a = \sum_{l=m}^{\infty} \left\{ \frac{1}{r}
        {\cblue W_l} Y_l^m \nabla^a r + {\cblue V_l} \nabla^a Y_l^m 
        - i {\cred U_l} \epsilon^{abc} \nabla_b Y_l^m \nabla_c r 
        \right\} e^{i \sigma t},
\label{v_exp}
\end{equation}
and solve the order $\Omega$ perturbation equations for the 
coefficients $\cred U_l(r)$ of the axial-parity terms and the 
coefficients $\cblue W_l(r)$ and $\cblue V_l(r)$ of the polar-parity 
terms. These coefficients are shown in Fig. \ref{fig:newt} for a 
particular hybrid mode of a uniform density Newtonian model.  For a 
pure r-mode, only one of the coefficients $\cred U_l(r)$ would be
nonzero.

R-modes of rapidly (and differentially) rotating Newtonian 
polytropes have been recently computed by Karino et al.~\cite{kyye00}
and references to the earlier Newtonian literature are given 
in the Introduction's bibliography above.  

\subsection{r-Modes of Relativistic Stars}

The r-modes of rotating relativistic stars were studied for the first
time only recently.\cite{a97,bk99,k98,kh99,kh00,lock99,laf00,rk01,sf00,y01}
As in the Newtonian case,\cite{lsvh92,lf99} a spherical barotropic 
relativistic star has a large degenerate subspace of zero-frequency
modes consisting of the axial-parity r-modes and the polar-parity
g-modes.  Although barotropic Newtonian stars retain a vestigial set of
purely axial $l=m$ modes, {\it\cdm rotating relativistic
stars of this type have  no  pure r-modes},\cite{lock99,laf00}  no
modes whose limit for a spherical star is purely axial.  Instead, the
Newtonian r-modes with $\cblue l=m\geq 2$ acquire relativistic
corrections with both axial and polar parity to become discrete
rotational 
modes of the corresponding relativistic models.  

\begin{figure}
\centerline{\psfig{file=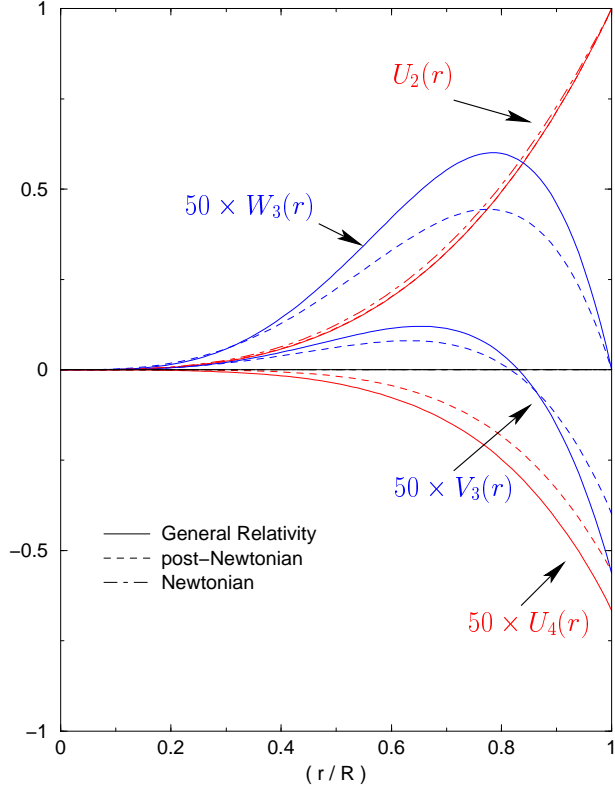,height=4.25in}}
\caption{Relativistic corrections to the $l=m=2$ r-mode of a uniform
density star of compactness $2M/R=0.316$. In Newtonian gravity, only
the coefficient $\cred U_2(r)$ is nonzero, i.e. the mode is a pure 
r-mode. However, the relativistic corrections turn the mode into 
a hybrid.}
\label{fig:grmode}
\end{figure}

Relativistic modes have been computed analytically in a post-Newtonian 
approximation\cite{laf00} and numerically for slowly rotating 
polytropes.\cite{laf01,rk01,y01} (See Fig. \ref{fig:grmode}.)
They have also been studied in numerical time evolutions of rapidly 
rotating relativistic stars using the Cowling approximation (hydro 
evolution with a frozen spacetime metric).\cite{sf00} Preliminary
results 
suggest that the growth timescale of the most unstable mode is largely 
unaffected by the relativistic corrections,\cite{laf01} differing from 
the post-Newtonian estimates\cite{lom98,aks98} by $\lesssim 10\%$.

In the slow-rotation approximation in which they have so far been
studied, axial perturbations of non-barotropic stars include, 
remarkably, a {\it continuous} spectrum.  Kojima~\cite{k98} shows that 
the axial modes are described by a single, second-order ODE for the
modes' radial behavior.  (See also the contributions of Kojima and
Hosonuma to the MG9 r-mode session, APT7).  He argues that the continuous 
spectrum is
implied by the vanishing of the coefficient of the highest derivative
term of this equation at some value of the radial coordinate, and Beyer
and Kokkotas\cite{bk99} make the claim precise.  As the latter authors
point out, the continuous spectrum they find may be an artifact of the
vanishing of the imaginary part of the frequency in the slow rotation
limit.  (Or, more broadly, it may be an artifact of the slow rotation
approximation.)

Furthermore, it is possible find discrete normal modes for
non-barotropic 
models. This is straightforward in the case of models for which the 
coefficient in the Kojima eqation is nonzero within the fluid.
(Non-barotropic uniform density models have this property and their 
r-modes have been computed by a number of
groups.)\cite{a98,laf00,rk01,y01} 
On the other hand, it is not so straightforward when the coefficient 
of Kojima's equation vanishes within the fluid - as is the case for
certain polytropes with realistic compactness.  Ruoff and 
Kokkotas\cite{rk01} and Yoshida\cite{y01} have argued that a discrete 
r-mode simply does not exist for such models.  However, it 
would be surprising if a small change in, say, the compactness of the 
star could lead to such a drastic change in the star's physics 
(the disappearance of the r-modes).  
It is perhaps more likely that the slow-rotation approximation is 
breaking down in the vicinity of the singular point and that if one 
were to regularize the Kojima equation in this region by including 
terms that are higher order in $\Omega$, one would be able to find 
discrete r-mode solutions.\cite{a98,laf01}

\section{Spin-down and Gravitational Radiation}

Early work suggested that the r-mode instability may sharply limit the
spin of newly formed, rapidly rotating neutron stars.  The radiation
may carry off almost all of the star's initial rotational energy
amounting to few percent of its rest mass, and the waves may be
detectable by LIGO II\cite{lom98,aks98,o98,bc98} from events out to 20 Mpc.  The spin-down
model on which these tantalizing estimates were based assumed that the
most unstable r-mode (with multipole indices $\cblue l=m=2$) would be able to
grow to an amplitude of order unity before being saturated by some sort
of nonlinear process, and, as we will see, that assumption has been
supported by subsequent work.

Spin-down computations in the context of linear perturbation theory
describe a competition between viscosity and gravitational radiation.
In these calculations,\cite{lom98,o98,aks98,lmo99} the growth time
$\cblue \tau$ of an unstable mode (or the damping time of a mode stabilized by
a viscosity) has the form,

\begin{equation}\cblue   \frac1\tau =  \color{dm} \frac1{\tau_{\rm GR}}\cblue + 
\frac1{\tau_{\rm shear viscosity}} +\frac1{\tau_{\rm bulk viscosity}},
\end{equation}
with $\cblue \tau$ the e-folding time for each process. (The analysis 
sketched below can be found in Lindblom et al\cite{lom98}; see also
Ipser and Lindblom\cite{il91}). When the energy radiated per cycle 
is small compared to the energy 
of the mode, the imaginary part of the mode frequency is accurately 
approximated by the expression
\begin{equation}\cblue \frac{1}{\tau} = - \frac{1}{2E} \frac{dE}{dt},   
\label{tau}
\end{equation}
where $\cblue E$ is the energy of the mode as measured in the rotating frame,
\begin{equation}\cblue E = \frac{1}{2} \int \left[ 
\rho \delta v^a \delta v^{\ast}_a 
+ \left( \frac{\delta p}{\rho}+\delta\Phi\right) 
\delta\rho^{\ast}
\right] d^3 x .
\label{E}
\end{equation}
We have,
\begin{eqnarray}\cblue 
\frac{dE}{dt} &=& -\color{dm} \sigma (\sigma + m\Omega) \sum_{l\geq 2} \cb N_l
\color{dm}\sigma^{2l} \left(
\left|\delta D_{lm}\right|^2 + \left|\delta J_{lm}\right|^2
\right) \nonumber \\
 & & \cblue - \int \left( 
2\eta\delta\sigma^{ab}\delta\sigma_{ab}^{\ast} 
+\zeta \delta\theta\delta\theta^{\ast}
\right), 
\label{dEdt}
\end{eqnarray}
where the dissipation due to gravitational radiation~\cite{th80} has coupling constant
\begin{equation}\cblue N_l = \frac{4\pi G}{c^{2l+1}}\frac{(l+1)(l+2)}{l(l-1)[(2l+1)!!]^2};
\end{equation}
$\cblue\delta\sigma_{ab}$ and $\cblue\delta\theta$ are the coefficients of shear
and bulk viscosity; and estimates\cite{cl92,s89} of corresponding coefficients $\cblue\eta$ and
$\cblue\zeta$ are 
\begin{equation}\cblue\eta = 2\times 10^{18} 
\left(\frac{\rho}{10^{15}\mbox{g}\!\cdot\!\mbox{cm}^{-3}}\right)^{\frac{9}{4}}
\left(\frac{10^9K}{T}\right)^2 \ 
\mbox{g}\!\cdot\!\mbox{cm}^{-1}\!\cdot\!\mbox{s}^{-1},
\label{eta}
\end{equation}
and 
\begin{equation}\cblue\zeta = 6\times 10^{25} 
\left(\frac{1\mbox{Hz}}{\sigma + m\Omega}\right)^2
\left(\frac{\rho}{10^{15}\mbox{g}\!\cdot\!\mbox{cm}^{-3}}\right)^2
\left(\frac{T}{10^9K}\right)^6 \ 
\mbox{g}\!\cdot\!\mbox{cm}^{-1}\!\cdot\!\mbox{s}^{-1};
\label{zeta}
\end{equation}

Polar and axial radiation arise, respectively, from mass and current
multipoles, $\cdm D_{lm}$ and $\cdm J_{lm}$, given by the equations,
\begin{equation}\cdm 
  D_{lm}  = \int   dV r^l  \rho   Y_{lm}^*\qquad\qquad   \ \ \ 
  J_{lm}  = {\cb\frac{2}{(l+1)c}}\int  dV r^l  \rho {
\bf v  \cdot r} \times\nabla Y_{lm}^*
\end{equation}
The additional factor of $\cdm v$ in the current multipoles implies an
additional factor of $\cdm v^2$ in the radiated energy of axial modes, 
and hence a smaller expected rate of radiation for the same multipole: 
For a mode of amplitude $\cblue A$  = (\cblue displacement of fluid element\cb)/$\cblue R$\cb,
with $\cblue R$ the stellar radius, we have  
\begin{equation}\cblue 
         \frac{dE}{dt}  \sim A^2  M^2 \cdm R^{2l} \sigma^{2l+2}\qquad\qquad 
  \cblue \frac{dE}{dt}  \sim A^2  M^2 \cdm R^{2l +2} \sigma^{2l +4}. 
\end{equation}

The extra factor of $\cblue \sigma^2$ in $\cblue dE/dt$ for the current multipoles
would make polar modes dominant if it were not for the fact that their
frequencies are small when they are unstable.  If a newborn neutron star
rotates at nearly its maximum frequency it is likely initially to be
unstable to both polar and axial modes.  Most of its spin-down,
however, should be dominated by the $\cblue l=m=2$ r-mode.  An optimistic
diagram of this spin-down is shown below in Fig. \ref{fig:owen}. (This figure, prepared by Ben
Owen, revises a similar figure in Lindblom, Owen and Morsink
\cite{lom98})  The perturbation here is assumed to reach saturation
with an amplitude of order unity, while still maintaining the character
of a linear r-mode.

The solid (purple) curve shows, for each temperature $\cblue T$, the minimum angular
velocity above which gravitational radiation can dominate viscosity and
drive an r-mode instability.  Below the solid curve on the
right, bulk viscosity due to neutrino production in URCA reactions
damps the instability.  Below the same solid curve on the left, shear
viscosity damps the instability. The upper red curves exhibit the   
larger shear viscosity from a laminar boundary layer when a crust is 
present (see Sec. \ref{sec:crust} below). A newborn star that starts with
angular velocity $\cblue \Omega_K$ follows the dashed trajectory when no crust
is present, becoming unstable when the temperature drops
below about $3\times 10^{10}K$.  The mode quickly reaches its saturation
amplitude and then radiates the star's angular momentum in
gravitational waves.  Finally, as the star cools and spins down, shear
viscosity damps the instability.   In the diagram, this occurs at a
temperature of about $10^9 K$ (after about a year, with standard
cooling), and at an angular velocity below $0.2\Omega_K$.  
As noted in Sect. ~\ref{nonlinear}, the amplitude may grow large 
enough that the later part of the nonlinear evolution no longer resembles 
this linear model.

\begin{figure}
\centerline{\psfig{file=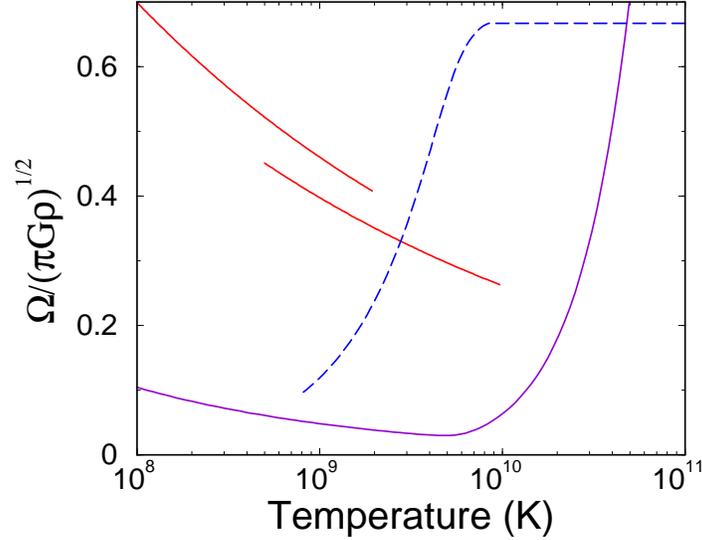,height=3.0in}}
\caption{Critical angular velocity vs. temperature for an $n=1.0$ polytrope.
Above the solid (purple) curve, the star rotates rapidly enough for its 
fastest growing ($\cblue l=m=2$) r-mode to be unstable, whereas below the curve all
modes are damped by viscosity.  The dashed curve shows the
evolution of a rapidly rotating neutron star as it cools and spins down
due to the emission of gravitational waves. The red curves show 
the larger critical angular velocities required for instability when 
a crust is present, the upper red curve corresponding to the viscosity 
when a superfluid is present. }
\label{fig:owen}
\end{figure}

This initial scenario is likely to overestimate the duration of the 
instability and the amount of energy radiated, and we discuss below
the principal corrections that have been considered.

\begin{figure}
\centerline{\psfig{file=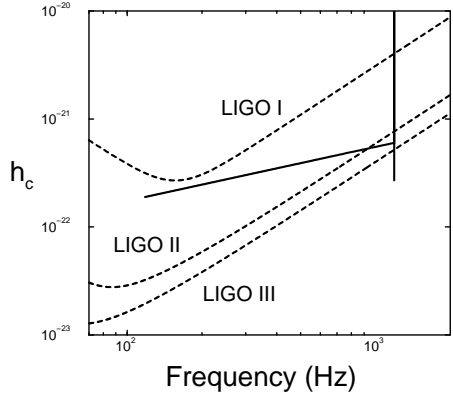,height=2.0in}}
\vskip 0.3cm
\caption{Characteristic gravitational wave amplitude $h_c$ (solid curve)
compared to the noise amplitude $h_{\rm rms}$ (dashed curves) for LIGO,
assuming perfect knowledge of the r-mode waveform.}
\label{fig:wave}
\end{figure}

If the r-modes saturate near unity, a perfect understanding of their
waveform would allow a high signal to noise ratio for sources beyond 
20 Mpc for detectors with the sensitivity of LIGO II, as dramatized 
by Fig. \ref{fig:wave} (taken from Owen et al.~\cite{o98} with permission
of the author).  

In reality, however, the large uncertainty in the waveform
substantially reduces the prospects for detection in the immediate
future.  A study by Brady and T.~Creighton~\cite{bc98}, following the work
by Owen et al\cite{o98}, sets a lower limit on detectability by
assuming no knowledge of the source, finding that newly formed neutron
stars should be detectable by LIGO II with narrow banding out to about
8 Mpc, with uncertainty allowing a range of perhaps 4-20 Mpc.  The
Virgo cluster would then likely be out of reach, and r-modes could not
be detected until more sensitive detectors were available.  

A stochastic background of gravitational waves produced by a cosmological
population of newly formed neutron stars is also likely to be detected
only by interferometers with advanced sensitivities (Owen et
al.~\cite{o98}, Ferrari et al.~\cite{fms98}, Schneider et
al.~\cite{sfm99}).

\subsection{Role of a Solid Crust}
\label{sec:crust}

Beginning with Bildsten and Ushormirsky,\cite{bu00} several authors
have considered a large increase in shear viscosity in a boundary layer
near a solid crust.\cite{ajks00,r00,lu00,lou00,wma00,m01}  Because the melting
temperature of the crust is estimated to be $10^{10}$K, with a factor
2 uncertainty, a crust may alter the r-mode instability
of newly formed stars and will certainly be important in any r-mode
instability of old stars spun up by accretion.

If the mode's velocity field vanishes in the crust, it will fall
rapidly to zero in a boundary layer.  In laminar flow,\cite{ll} the
thickness of the boundary layer ({\it Eckman} layer),
\be \cdm
        d = \sqrt{\frac\nu{2\Omega}},
\ee
can be estimated by equating the acceleration 
$\cblue \delta [(\partial_t v\cdot)\nabla v)] \sim \Omega\delta v$
of a fluid element to the viscous force per unit mass, $\cblue \sim \eta \delta
v/\rho d^2$.  For a neutron star above the superfluid transition
temperature, the coefficient of viscosity $\cblue \eta$ is of order $\cblue \eta =
2\times 10^{18} \rho_{15}^{9/4} T_9^{-2}$ gm/cm-s, and $\cblue d$ is a few cm.

As this rough estimate suggests, when a crust is present, dissipation
in the viscous boundary layer dominates viscous energy loss in Eq.
(\ref{dEdt}):  Because the integrand
$\cblue \eta\delta\sigma^{ab}\delta\sigma_{ab}$ in the layer is enhanced by a
factor of order $\cblue R^2/d^2$ ($\cblue R$ the radius of the star), the dissipation
in the layer is larger than that in the interior by a factor $\cblue R/d$.
Estimate of the higher angular velocities needed for instability 
are shown as the red curves in Fig. ~\ref{fig:owen}.

Levin and Ushomirsky\cite{lu00} point out that this conclusion somewhat
overstates the case, because the crust will participate in an r-mode.
Detailed calculations by Yoshida and Lee~\cite{yl01} describe the
interaction of the stellar r-modes with r-modes of the crust, finding a
series of avoided crossings.  Levin and Ushomirsky find that the
avoided crossings lead to a wide variation in the fractional drop in
$\cblue \delta v$ across the boundary layer, which can range from 1/20 to 1,
allowing a possible decrease in the boundary layer dissipation by a
factor $(1/20)^2 = 1/400$, compared to the estimate above.

These estimates are based on an assumption of laminar flow, but 
Wu et al\cite{wma00} show that in a maximally rotating star, 
a mode whose amplitude exceeds $\sim 10^{-3}$  will become turbulent.
Because the relevant amplitudes are larger than this, theirs is  
the best current computation of viscous damping in the presence 
of a crust.

The high boundary-layer viscosity may not significantly alter an 
r-mode instability of newborn neutron stars, because the $\cblue l=m=2$ r-mode
is likely to be unstable by the time the temperature has dropped to
between $2\times 10^{10}K$ and $4\times 10^{10}K$, slightly above the
temperature at which the crust is expected to solidify.  If this is the
case, Lindblom, Owen, and Ushomirsky\cite{lou00} show that the r-mode will
prevent a solid crust from forming.  Instead, they argue, chunks of ice
will lie in the outer part of star,  with a density adjusted to make
the heat dissipated by the ice flow balance the mode's
gravitational-wave driven rate of energy increase.  If, as expected,
the mode amplitude is above $10^{-4}$ before a crust forms, the ice
flow will apparently not significantly affect the r-mode spin-down.

\subsection{Instability in Old Stars Spun-Up by Accretion?}
\label{sec:accretion}

For old neutron stars spun up by accretion, the crust is
likely to play a crucial role.  Wagoner's hope\cite{w84} was that
accreting neutron stars would spin up until they were unstable (to
polar modes, in his paper) and would then radiate angular momentum in
gravitational waves at a rate that balanced the angular momentum gained
in accretion.  Bildsten~\cite{b98} and Andersson, Kokkotas, and 
Stergioulas~\cite{akst98} considered the mechanism for r-modes in 
low-mass x-ray binaries (LMXBs), the latter suggesting that it might account
for the narrow range of observed angular velocities.    

Brown and Ushomirsky~\cite{bru00}, however, concluded that low observed
luminosities were inconsistent with the Wagoner mechanism.
Levin,~\cite{l98} reexamined the mechanism for r-modes, finding a
cycle in which the star is spun up to instability, but reaches no
equilibrium.  Instead, by heating up the star, the mode lowers the
viscous damping, becomes increasingly unstable, and radiates its energy
in a short time (less than a year).  It is then gradually spun up (for
more than $10^6$ years), and the cycle repeats.  Armed with this
revised scenario, Andersson et al\cite{ajks00} suggested that the
instability is responsible for the maximum observed angular velocity
(1.6 ms) of old neutron stars and is consistent with the observed range
of LMXB spins.

Despite the large difference between turbulent boundary-layer viscosity
and the shear viscosity of the neutron-star interior that Andersson et
al.~\cite{ajks00} had used, Wu et al.\cite{wma00} find similar results; that is, like Andersson
et
al.~\cite{ajks00}, they find spin-rates consistent with those of LMXBs.  The maximum
angular velocity grows with the accretion rate and falls in the narrow
range of 490-700 Hz for $\cblue \dot M/M$ between $10^{-11}$ and
$10^{-8}\ {\rm yr}^{-1}$, when the crust is rigid.  (If the fractional
drop in $\cblue \delta v$ across the boundary layer is only $1/10$, the
maximum angular velocities fall to 220 - 300 Hz; and an unexpectedly
low melting temperature for the crust will again reduce the maximum
$\cblue \Omega$.)

Levin's cycle is qualitatively similar in this analysis, but the spin-down
time increases to $10^3$ yr, and the spin-up time to $10^7$ yr. 

In a study of hypercritical accretion flow onto neutron stars, Yoshida and 
Eriguchi~\cite{ye99} find that an unstable r-mode strongly limits the neutron-star spin.
The high energy input associated with accretion is large enough that Levin's
cycle is not seen.  These are then a class of systems, possible precursors
to compact binary systems, in which accreting neutron stars may reach 
a Wagoner equilibrium.   

Finally, we should mention a recent preprint by Mendell,~\cite{m01} examining the 
damping of r-modes by a magnetic field in a viscous boundary layer.
The damping is large for magnetic fields of order $10^{12} G$, and, 
like the boundary-layer viscosity, could prevent an r-mode instablity 
in a newborn star if a crust forms before an unstable mode has time to grow. 

\section{Nonlinear Calculations} 
\label{nonlinear}

Much of the very recent work on the r-mode instability has addressed the
nonlinear evolution of the r-modes.  The central issue is whether the instability found in
idealized
models survives the physics that governs a young neutron star.
We discussed the role of the crust in the last section, and now
ask two questions related to a mode's nonlinear evolution:
Does nonlinear coupling to other modes allow an unstable r-mode
to grow to unit amplitude? Does the background star retain a uniform
rotation law as it spins down or does a growing r-mode generate
significant differential rotation?
The importance of this last question was emphasized by Spruit
\cite{s99} and by Rezzolla, Lamb and Shapiro~\cite{rls99} who argued
that  differential rotation would wind up a toroidal magnetic field
and drain the oscillation energy of the r-mode.  (See also Rezzolla's
contribution to the MG9 r-mode workshop APT7).
A number of different approaches have since been applied to the
nonlinear r-mode problem in an attempt to address these questions.

One notable approach is the direct numerical evolution of the
nonlinear equations describing a self-gravitating fluid.  Stergioulas
and Font~\cite{sf00} have performed 3-D general relativistic hydrodynamic
evolutions in the Cowling approximation, and Lindblom, Tohline
and Vallisneri~\cite{ltv00} have performed 3-D Newtonian hydrodynamic
evolutions with an added driving force representing gravitational
radiation-reaction, equivalent to that computed previously by Rezzolla 
et al.~\cite{rea99}

Stergioulas and Font~\cite{sf00} construct an equilibrium model of a
rapidly rotating relativistic star and add to it an initial
perturbation that roughly approximates its $\cblue l=m=2$ r-mode.
They then evolve the perturbed star using the nonlinear hydrodynamic
equations with the spacetime metric held fixed to its equilibrium value
(the relativistic Cowling approximation).
They find no evidence for suppression of the mode on a dynamical
timescale, even when the mode amplitude, $\cblue A$, is initially
taken to be of order unity.
Because of the approximate nature of the initial perturbation, other
oscillation modes are excited in the initial data. Recall that for a star with a
barotropic equation of state, the generic rotationally restored
mode is not a pure axial-parity r-mode, but an r-g ``hybrid'' mode (Sect.
\ref{sec:mode_character}).
Stergioulas and Font~\cite{sf00} find that a number of these hybrid
modes are excited in their initial data with good agreement between
the inferred frequencies and earlier results from linear perturbation
theory~\cite{lf99}. In their published work, they find no evidence
that the dominant mode is leaking its oscillation energy to other
modes on a dynamical timescale.  Instead, a nonlinear version of an
r-mode appears to persist over the time of the run, about 25 rotations
of the star. In additional runs with amplitudes substantially larger
than unity, however, one no longer sees a coherent r-mode.  This may
be evidence of nonlinear saturation, but further runs with more accurate
initial data will be necessary to conclude this definitively~\cite{s01}.
(See also Stergioulas' contribution to the r-mode workshop APT7).

These conclusions are consistent with preliminary results from studies
of nonlinear mode-mode couplings at higher order in perturbation
theory~\cite{saftw01,m00}.
Other r-modes of a nonbarotropic star seem to give no
indication of a strong coupling to the $\cblue l=m=2$ r-mode unless its
amplitude is unphysically large ($\cblue A\grsim 30$!)~\cite{m00}.  Work is
still in progress on the nonlinear coupling of the dominant r-mode to
the g-modes of nonbarotropic stars~\cite{m00} and to the hybrid modes of
barotropic stars~\cite{saftw01}.

The results of Stergioulas and Font~\cite{sf00} have also been confirmed and
significantly extended by the calculation of Lindblom, Tohline and
Vallisneri.~\cite{ltv00} In Stergioulas and Font's calculation the growth
of the unstable r-mode does not occur because the spacetime dynamics have
been turned off.  However, it would be impossible to model this growth
anyway even in a fully general relativistic hydrodynamic evolution,
because the timescale on which the mode grows due to the emission of
gravitational waves far exceeds the dynamical timescale of a rapidly
rotating neutron star.

To simulate the growth of the dominant r-mode in a calculation
accessible to current supercomputers, Lindblom, Tohline and Vallisneri
\cite{ltv00} take a different approach. They begin by constructing an
equilibrium model of a rapidly rotating Newtonian star and add to
it a small initial perturbation corresponding to its $\cblue l=m=2$ r-mode.
They then evolve the perturbed star by the equations of Newtonian
hydrodynamics with a post-Newtonian radiation-reaction force that
drives the current quadrupole associated with the $\cblue l=m=2$ r-mode.

By artificially scaling up the strength of the driving force, they
are able to shorten the growth time of the unstable r-mode by a factor
of $4500$.  In the resulting simulation the mode grows exponentially
from an amplitude $\cblue A=0.1$ to $\cblue A=2.0$ in only about 20
rotations of the star.

With this magnified radiation-reaction force, Lindblom, Tohline and
Vallisneri~\cite{ltv00} are able to confirm the general features of the
simplified r-mode spin-down models~\cite{lom98,aks98,o98,bc98}. In their simulation,
the star begins to spin down noticeably when the amplitude of the dominant
mode is of order unity, and ultimately about $40\%$ of the star's
angular momentum is radiated away. The evolution of the star's
angular momentum as computed numerically agrees well with the predicted
angular momentum loss to gravitational radiation.
If their model is accurate, however, gravitational radiation would
not be emitted steadily at a saturation amplitude, but would die out
after saturation and then reappear as the mode regenerates.

Again, there is no evidence of nonlinear saturation for mode amplitudes
$\cblue A\lesssim 1$.  The growth of the mode is eventually suppressed
at an amplitude $\cblue A\simeq 3.4$, and the amplitude drops off sharply
thereafter.  Lindblom, Tohline and Vallisneri argue that the mechanism
suppressing the mode is the formation of shocks associated with the
breaking of surface waves on the star.  They find no evidence of
mass-shedding, nor of coupling of the dominant mode to the other
r-modes or hybrid modes of their Newtonian barotropic model.

These various studies all provide evidence pointing to the same
conclusion: the most unstable r-mode appears likely to grow to an
amplitude of order unity before being suppressed by nonlinear
hydrodynamic processes. It is important to emphasize, however,
that the 3-D numerical simulations have probed nonlinear processes
occurring only on dynamical timescales and that the actual growth
timescale for the r-mode instability is longer by a factor of order $10^4$.
It is possible that the instability may be suppressed by hydrodynamic
couplings occurring on timescales that are longer than the dynamical
timescale but shorter than the r-mode growth timescale.
Further work clearly needs to be done before definitive conclusions
can be drawn. Particularly relevant will be the results from the
ongoing mode-mode coupling studies~\cite{saftw01,m00}.

Turning to the question of differential rotation, deviations from
a uniform rotation law are observed in both of the 3-D numerical
simulations~\cite{sf00,ltv00}.
It has been proposed that differential rotation will be driven by
gravitational radiation-reaction~\cite{s99} as well as being associated
with the second order motion of the r-mode, itself~\cite{rls99}.
In a useful toy model, Levin and Ushomirsky~\cite{lu99} calculated an
exact r-mode solution in a thin fluid shell and found both sources
of differential rotation to be present.

To address in more detail the issue of whether or not the r-mode
instability would generate significant differential rotation,
Friedman, Lockitch and S\'a~\cite{fls01} have calculated the
axisymmetric part of the second order r-mode.
We work to second order in perturbation theory with the equilibrium
solution taken to be either a slowly rotating polytrope (with index
$\cblue n=1$) or an arbitrarily rotating uniform density star (a Maclaurin
spheroid).  The first order solution, which appears in the source
term of the second order equations, is taken to be a pure $\cblue l=m$
r-mode with amplitude $\cblue A$.

We find that differential rotation is indeed generated both by
gravitational radiation-reaction and by the quadratic source terms
in Euler's equation; however, the latter dominate a post-Newtonian
expansion.
The functional form of the differential rotation is independent
of the equation of state - the axisymmetric, second order change
in $\cblue v^\varphi$ being proportional to $\cblue z^2$ (in cylindrical coordinates)
for both the polytrope and Maclaurin.

Our result extends that of Rezzolla, Lamb and Shapiro~\cite{rls99}
who computed the order $\cblue A^2$ differential drift resulting from
the linear r-mode velocity field.
These authors neglect the nonlinear terms in the fluid equations
and argue (based on an analogy with shallow water waves) that the
contribution from the neglected terms might be irrelevant.  Indeed,
for sound waves and shallow water waves, the fluid drift computed using
the linear velocity field turns out to be exact to second order
\cite{lmrs99}; thus, one may safely ignore the nonlinear terms.
However, for the motion of a fluid element associated with the r-modes,
we find that there is in fact a non-negligible contribution from the
second-order change in $\cblue v^\varphi$. Interestingly, the resulting second
order differential rotation is stratified on cylinders.
It remains to be seen whether the coupling of this differential rotation
to the star's magnetic field does indeed imply suppression of the r-mode
instability.

\section{Comments on Future Work}

The overarching questions are whether unstable r-modes limit the spin of either young, rapidly
rotating neutron stars or of old stars spun up by accretion; and if young stars are unstable,
how accurately can 
we characterize the r-mode's wave form?  

The nonlinear studies of r-mode instability are still at an early
stage, and it is not yet certain what limits the amplitude of an
r-mode. All computations so far agree that there is no saturation until
large amplitude is reached, and both numerical groups observe the
formation of shocks and surface-wave breaking at large
amplitudes.~\cite{ltv00,s01}  Still unclear, however, is whether sudden 
damping of the mode by wave breaking is an artifact of the 
numerics or of an amplified radiation reaction; the possibility 
remains that mode-mode coupling enforces a smoother limit 
on the r-mode amplitude.  

It is similarly unclear whether an r-mode will wind up a magnetic 
field or whether, in the presence of a background magnetic field,
there is a slightly altered mode that does not secularly
change the field.   

On the mathematical side, deciding whether the spectrum of rotational modes 
of relativistic stars is continuous or discrete may be a tractable 
problem and may have implications for neutron-star physics.

More generally, uncertainties in the microphysics (in, e.g., bulk viscosity
due to hyperon production in the core, or in the coupling of superfluid 
and crust) are large enough that we are unlikely to decide soon whether 
unstable r-modes play a role in the lives of neutron stars.  But a 
discovery of their gravitational waves could decide the issue for us.  
 
\section*{Acknowledgments}
We thank Ben Owen for helpful conversations.  Work to prepare this
review was supported in part by the National Science Foundation under
Grants PHY00-71044 and PHY95-14240 and by the Eberly research funds of
Penn State.

\end{document}